\documentclass[12pt,titlepage]{article}
\usepackage{bm, amssymb,pifont,cancel, amsmath,comment,color}
\usepackage[dvips]{graphicx}
\makeatletter

\@addtoreset{equation}{section}
\makeatother
 
\begin{document}

\begin{titlepage}
\null
\begin{flushright}
WU-HEP-14-05
\end{flushright}

\vskip 1.5cm
\begin{center}
\baselineskip 0.8cm
{\LARGE \bf
Illustrating SUSY breaking effects on various inflation mechanisms}

\lineskip .75em
\vskip 1.5cm

\normalsize

{\large Hiroyuki Abe} $\!${\def\thefootnote{\fnsymbol{footnote}}\footnote[1]{E-mail address: abe@waseda.jp}}, 
{\large Shuntaro Aoki} $\!${\def\thefootnote{\fnsymbol{footnote}}\footnote[2]{E-mail address: shun-soccer@akane.waseda.jp}}, 
{\large Fuminori Hasegawa} $\!${\def\thefootnote{\fnsymbol{footnote}}\footnote[3]{E-mail address: corni\_in\_f@akane.waseda.jp}},

 {\large and} {\large Yusuke Yamada} $\!${\def\thefootnote{\fnsymbol{footnote}}\footnote[4]{E-mail address: yuusuke-yamada@asagi.waseda.jp}}

\vskip 1.0em

{\small\it Department of Physics, Waseda University, \\ 
Tokyo 169-8555, Japan}

\vspace{12mm}

{\bf Abstract}\\[5mm]
{\parbox{13cm}{\hspace{5mm} \small
We consider the supersymmetry breaking effects on typical inflation models with different types of K\"ahler potential. The critical size of supersymmetry-breaking scale, above which the flatness of the inflaton potential is spoiled, drastically changes model by model.  We present the universal description of such effects in terms of a field-dependent scaling factor by which gravity-mediated supersymmetry breaking terms are suppressed or enhanced, based on the conformal supergravity framework. Such a description would be useful when we estimate them in constructing supersymmetric models of particle cosmology. 
}}

\end{center}

\end{titlepage}

\tableofcontents
\vspace{35pt}
\hrule
\section{Introduction}
Cosmic inflation~\cite{Guth:1980zm}  is the most promising candidate for a solution to the flatness and horizon problem in the standard cosmology, which also explains the existence of the primordial density fluctuation. Especially, slow-roll inflation models naturally realize the almost scale invariant fluctuation of the curvature, and the predicted spectrum of the fluctuation is tested by the observation of the cosmic microwave background~(CMB). 

The observational results from Planck satellite~\cite{Ade:2013uln} show the small tensor-to-scalar ratio~$r$ and the spectral tilt $n_s\sim 0.96$, which can be naturally realized in the Starobinsky model~\cite{Starobinsky:1980te} and the Higgs inflation model with a non-minimal coupling between the Higgs boson and the Ricci scalar~\cite{Bezrukov:2007ep}.
On the other hand, the recent discovery of the primordial gravitational wave reported by BICEP2 experiment~\cite{Ade:2014xna} strictly constrains many inflation models if it is verified by the other experiments. The result implies that the field value of the inflaton during inflation will be much larger than the Planck scale~$M_{\rm pl}\sim 2.4\times 10^{18}$GeV, e.g. the simple chaotic inflation~\cite{Linde:1983gd}, and then estimating the effects of Planck-suppressed terms in the scalar potential become more and more important in such a model building.

In supergravity-based inflation models, such Planck-suppressed terms arise through the K\"ahler potential in general.  In most cases, such terms spoil the flatness of the inflaton potential, that is, the so-called $\eta$-problem arises. Therefore, it is important to impose certain symmetries to K\"ahler potential terms which protect the enough flatness of the potential. For example, realization of the chaotic inflation in supergravity~\cite{Kawasaki:2000yn} requires a shift symmetry for the inflaton multiplet.  In Ref.~\cite{Roest:2013aoa}, to realize the single field inflation models in supergravity, more general K\"ahler potential structures including such as the Heisenberg symmetry are discussed.

The structure of the K\"ahler potential is also important from the viewpoint of the supersymmetry~(SUSY) breaking effects. SUSY should be broken at some scale higher than the electroweak scale. In many cases, it is assumed that SUSY is broken in a hidden sector, and the standard model sector feels its effects mediated by the gauge or Planck suppressed interactions typically. For the later case, the structure of the K\"ahler potential plays the crucial role to determine the structure of soft supersymmetry breaking terms relevant to the (super)particle phenomenology at a low energy. 

It is remarkable that the SUSY breaking effects required at the minimum of the scalar potential (where the present universe resides in) can in general affect the global structure of inflaton potential (the trajectory of inflaton dynamics toward its minimum). In some works (e.g. \cite{Nakayama:2010xf, Takahashi:2013cxa,Buchmuller:2014pla}), it was discussed that such effects may spoil the flatness of the inflationary trajectory through the Planck suppressed operator, which is mostly related to the structure of the K\"ahler potential.

In this paper, we will discuss the relation between the SUSY breaking effects at the minimum of the scalar potential and those during the inflation, by focussing on the inflaton K\"ahler potential. We will notice that the structure of the K\"ahler potential governs the behavior of terms induced by the SUSY breaking. We will show that the essential properties of such terms can be easily understood from a field-dependent rescaling factor of the physical mass scales based on the conformal supergravity framework~\cite{Kugo:1982cu}.

The remaining of this paper consists of four sections. In Sec.~2, we will discuss the SUSY breaking effects in three inflation models with different types of K\"ahler potential. Then we will find that the critical size of the SUSY breaking scale, above which the inflationary potential is spoiled, is totally different model by model depending on their K\"ahler structure. From the conformal supergravity point of view, we illustrate such a difference in terms of the rescaling factor of gravity-mediated SUSY breaking terms in the Einstein frame in Sec.~3. After that, we will focus on the class of models which are compatible with the recent results from the BICEP2 experiment in Sec.4. Finally, we conclude in Sec.~5.
\section{SUSY breaking effects on various inflation models}\label{models}
In this section, we discuss the SUSY breaking effects in three illustrative inflation models which contain different types of K\"ahler potential. In every model, for comparison, we assume the SUSY breaking sector $X$ with the same superpotential terms,
\begin{eqnarray}
W_{\cancel{\rm SUSY}}=W_0+\mu^2X,
\end{eqnarray} 
where $W_0$ and $\mu$ are complex constants in general but we assume they are real for simplicity because these complex phases do not play essential roles in our discussion.  Here and hereafter all the mass scales are measured in the unit $M_{\rm pl}=1$. 

Each of the three inflation models has a different one of the following three types of K\"ahler potential $K=K_{\rm min}, K_{\rm conf}, K_{\rm shift}$ from each other,
\begin{eqnarray}
K_{\rm min}&=&|\Phi|^2,\label{minimalK}\\
K_{\rm conf}&=&-3\log\left( -\frac{\Omega_{\rm conf}}{3}\right)=-3\log\left(-\frac{1}{3}(-3+|\Phi|^2+J(\Phi)+J(\bar{\Phi}))\right),\label{KCSS}\\
K_{\rm shift}&=& K_{\rm min}+\frac{1}{2}(\Phi^2+{\rm h.c.}) \ = \ \frac{1}{2}(\Phi+\bar{\Phi})^2,
\end{eqnarray}
where $\Phi$ is a chiral multiplet and $J(\Phi)$ is a holomorphic function of $\Phi$.
In the Einstein frame, the general supergravity~(SUGRA) Lagrangian is given by
\begin{eqnarray}
\mathcal{L}&=&-K_{I\bar{J}}\partial_\mu \phi^I\partial^\mu\bar{\phi}^{\bar{J}}-V_E+\cdots,\\
V_E&=&e^K\left[K^{I\bar{J}}(\partial_IW+\partial_IK W)(\partial_{\bar{J}}\bar{W}+\partial_{\bar{J}}K \bar{W})-3|W|^2\right]\nonumber\\
&&-\frac{1}{2}({\rm Re}~f_{AB})G_Ik_A^IG_Jk_B^I,\label{VE}
\end{eqnarray}
where $K_{I\bar{J}}=\partial _I\partial_{\bar{J}}K$ is the K\"ahler metric, $\phi^I$ is the scalar component of a chiral multiplet $\Phi^I$, $G=K+\log|W|^2$ denotes the K\"ahler invariant function, $k_I^A$ is the Killing vector of $\phi^I$ under a gauge group $A$, and the ellipsis denotes terms including higher-spin fields. Then  we find that both the following two types of K\"ahler potential $K_{\rm min}$ and $K_{\rm shift}$ give a canonical kinetic term for the chiral multiplet $\Phi$ contained in each of them.

On the other hand, the general form of the SUGRA Lagrangian in the Jordan frame~\cite{Ferrara:2010yw} is given by
\begin{eqnarray}
\mathcal{L}_{J}&=&-\frac{\Omega}{6} R-\left( \frac{1}{3}\Omega K_{I\bar{J}} -\frac{\Omega_I\Omega_{\bar{J}}}{\Omega}\right)\partial_\mu \phi^I\partial^\mu\bar{\phi}^{\bar{J}}-V_J+\cdots,\label{JL}\\
V_J&=&\frac{\Omega^2}{9}V_E,
\end{eqnarray}
where $V_E$ is shown in Eq.~(\ref{VE}) and the so-called frame function $\Omega$ is a real function of $\phi^I$ and $\bar{\phi}^{\bar{J}}$. In the case that $\Omega$ is related with the K\"ahler potential $K$ as 
\begin{eqnarray}
\Omega=-3e^{-\frac{1}{3}K},
\end{eqnarray}
we find a (non-)canonical kinetic term of $\phi^I$ for $K=K_{\rm conf}$ (for $K=K_{\rm min}$ or $K_{\rm shift}$) in this frame.
Therefore, each one of the three types of K\"ahler potential should be called ``minimal" in the different context. The first type, $K_{\rm min}$, and the third one, $K_{\rm shift}$, will be called minimal forms in the Einstein frame SUGRA, while the second one, $K_{\rm conf}$, should be referred to as the minimal one in such a sense that it produces a canonical kinetic term in Jordan frame SUGRA.\footnote{We should note that the Lagrangian~(\ref{JL}) contains a term $\delta\mathcal{L}=\mathcal{A}_\mu^2$, where $\mathcal{A}_\mu\equiv \frac{i}{6}(\partial_\mu \Phi^IK_I-\partial_\mu \bar{\Phi}^{\bar{J}}K_{\bar{J}})$, and therefore even if we choose the K\"ahler potential as $K_{\rm conf}$, the kinetic terms of a scalar is not a canonical form exactly. However, as discussed in Ref.~\cite{Ferrara:2010yw}, $\mathcal{A}_\mu$ vanishes on the inflationary trajectory in many inflation models, and then the kinetic term of the inflaton becomes a canonical one. In the later discussion, we discuss such kind of models.}

For the purpose of our study, we will employ the D-term hybrid inflation~\cite{Binetruy:1996xj}, the universal attractor inflation~\cite{Kallosh:2013tua}, and the chaotic inflation~\cite{Kawasaki:2000yn,Kallosh:2010ug}, each has the K\"ahler potential $K_{\rm min}$, $K_{\rm conf}$ and $K_{\rm shift}$ respectively, and study the SUSY breaking effects on these typical and illustrative inflation models\footnote{The first two inflation models with the original choice of parameters are not favored by the recent observational results if we combine both the data from Planck and BICEP2 experiments. We will discuss the models  with/without the different choice of parameters from the original ones, which are compatible with the BICEP2 result in Sec.4.}, looking at the difference between them.
\subsection{Minimal K\"ahler model}\label{mini}
First, we discuss a D-term hybrid inflation~\cite{Binetruy:1996xj} with $K_{\rm min}$. The K\"ahler potential and superpotntial are given by
\begin{eqnarray}
K&=&|\Phi|^2+|S_+|^2+|S_-|^2,\\
W&=&\lambda \Phi S_+S_-,
\end{eqnarray} 
where $\lambda$ is a real constant, $S_\pm$ are chiral multiplets with the charges $q_\pm=\pm1$ under a local U(1) symmetry, respectively.

As mentioned before, we simply add the following terms to consider the SUSY breaking effects on this model,
\begin{eqnarray}
\delta K&=&|X|^2-\frac{|X|^4}{\Lambda^2},\label{KX}\\
\delta W&=&W_{\cancel{\rm SUSY}},
\end{eqnarray} 
where $\Lambda\ (\ll1)$ is a real constant. 

The supersymmetric mass terms for $S_+$ and $S_-$ arise in the scalar potential if the VEV of $|S_\pm|$ is sufficiently large. Taking the condition into account that the cosmological constant vanishes at the minimum, the scalar potential can be expressed by
\begin{eqnarray}
V&\simeq& \frac{g^2\chi^2}{2}+e^{|\Phi|^2}|\Phi|^2W_0^2+m^2_+|S_+|^2+m_-^2|S_-|^2+m_X^2|X|^2\nonumber\\
&&+(m_{\pm}^2S_+S_-+\alpha_XX+{\rm h.c.}),
\end{eqnarray}
where $\chi$ is a so-called Fayet-Iliopoulos parameter in the D-term of local U(1) gauge multiplet, and
\begin{eqnarray}
m_+^2&=&e^{|\Phi|^2}[\lambda^2|\Phi|^2+(1+|\Phi|^2)W_0^2]+g^2\chi,\\
m_-^2&=&e^{|\Phi|^2}[\lambda^2|\Phi|^2+(1+|\Phi|^2)W_0^2]-g^2\chi,\\
m_\pm^2&=&e^{|\Phi|^2}(-1+|\Phi|^2)\lambda W_0\Phi.
\end{eqnarray}

If there are no SUSY breaking effects, $m_-^2$ is positive for $|\Phi|>\Phi_c$ where $\Phi_c$ satisfies the following equation,
\begin{eqnarray}
e^{|\Phi_c|^2}|\Phi_c|^2=\frac{g^2\chi}{\lambda^2}.
\end{eqnarray} 
When the VEV of $|\Phi|$ becomes smaller than $|\Phi_c|$, $S_-$ obtains the tachyonic mass and the inflation ends. We have to note that even if the SUSY breaking effects exist, the value of $|\Phi_c|$ is not so altered. 

After integrating out the heavy modes, we obtain the following effective potential for the inflaton field $|\Phi|$,
\begin{eqnarray}
V_{\rm eff}\simeq \frac{g^2\chi^2}{2}+e^{|\Phi|^2}|\Phi|^2W_0^2+\frac{g^4\chi^2}{16\pi^2}\left(1+\log\frac{\lambda e^{|\Phi|^2}|\Phi|^2}{Q^2}\right),\label{VD}
\end{eqnarray}
where the last term is produced by the one-loop effects and $Q$ is a renormalization scale which we choose $Q^2=g^2\chi$ here. 
\begin{figure}
\begin{center}
\includegraphics[width=\textwidth]{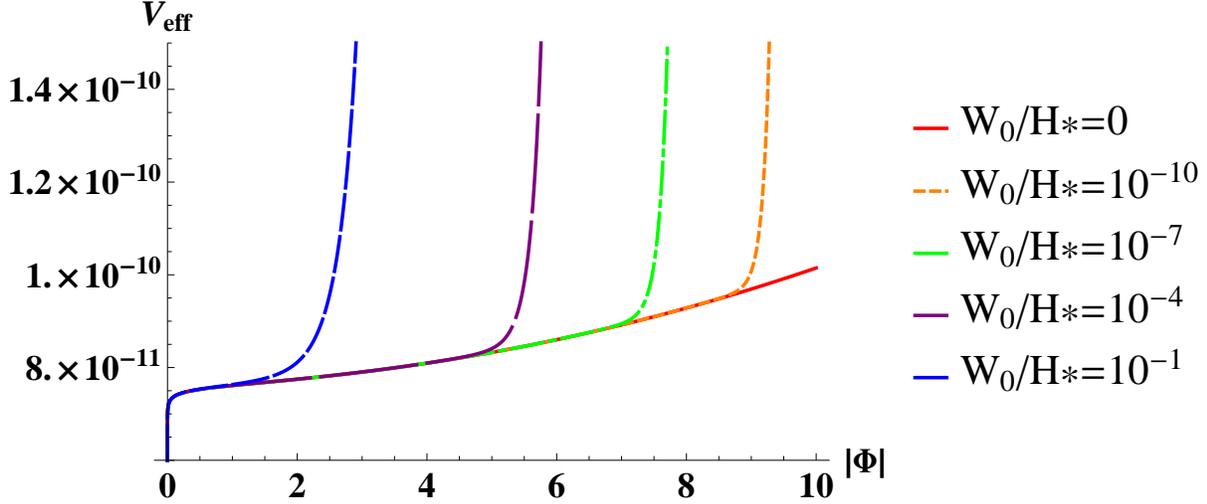}
\caption{The $|\Phi|$-dependences of the scalar potential~(\ref{VD}) with different values of $W_0$ are shown. The ratio between SUSY breaking scale $W_0\sim m_{3/2}$ and the inflation scale $H_*\equiv \sqrt{g^2\chi^2/6}$ is changed for each line.}
\label{Dterm_fig}
\end{center}
\end{figure}
In Fig.~\ref{Dterm_fig}, we show the behaviors of the effective potential with various values of $W_0\sim m_{3/2}$, where $m_{3/2}$ is the gravitino mass. The SUSY breaking scale increases as the parameter $W_0$ gets larger, and we find from Fig. 1, at the same time, the scalar potential becomes steeper in the large field region $|\Phi| \gtrsim\log_{10} \frac{H_*}{W_0}=  \log_{10} \frac{g\chi}{\sqrt{6}W_0}$. Therefore we can confirm that a successful inflation can not be realized if the ratio $W_0/H_*=\sqrt{6}W_0/g\chi$ is larger than $3\sqrt{3}\times 10^{-2}$.\footnote{Here, we choose the set of model parameters $(g, \chi, \lambda)$ as $(1/\sqrt{2},1.72\times 10^{-5}, 0.02)$.}
\subsection{The conformal model}\label{CSS}
In this subsection, we analyze the universal attractor inflation with SUSY breaking effects discussed in Ref.~\cite{Kallosh:2013tua}. The K\"ahler potential~$K$ and superpotential~$W$ are given by
\begin{eqnarray}
K&=&-3\log \left(-\frac{\Omega}{3}\right),\\
\Omega &=&\Omega_{\rm conf}+|S|^2-\frac{3\zeta}{2+\xi (f(\sqrt{2}\Phi)+f(\sqrt{2}\bar{\Phi}))}|S|^4,\label{Ocss}\\
J(\Phi)&=&-\frac{1}{2}\Phi^2-\frac{3}{2}\xi f(\sqrt{2}\Phi),\\
W&=&\lambda f(\sqrt{2}\Phi)S, 
\end{eqnarray}  
where $\xi$, $\lambda$ are real constants, and $\Phi$, $S$ are chiral multiplets. The inflaton field in this model is $\phi\equiv{\rm Re}~\Phi/\sqrt{2}$ and the other fields are stabilized at their origin during inflation. 

For $\xi\gg1$, independently to the choice of the function $f(\Phi)$, the spectral tilt of scalar perturbation $n_s$ and tensor-to-scalar ratio $r$ in this model are universally attracted to 
\begin{eqnarray}
n_s&=&1-\frac{2}{N_*},\\
r&=&\frac{12}{N_*^2},
\end{eqnarray}  
respectively, where $N_*$ is the number of the e-foldings before the inflation end. It is remarkable that this attractor point of $(n_s,r)$ is the central value of the Planck results~\cite{Ade:2013uln}.

We consider the SUSY breaking effects on this model by adding the following terms in $\Omega=-3e^{-K/3}$ and $W$,
\begin{eqnarray}
\delta\Omega&=&|X|^2-\frac{|X|^4}{\Lambda^2},\\
\delta W&=&W_{\cancel{\rm SUSY}}.
\end{eqnarray}

The scalar component of Goldstino multiplet $X$ (sGoldstino) can be stabilized at around its origin if the constant $\Lambda$ is much smaller than 1, as in the case of Sec.~\ref{mini}. Then, the effective potential during inflation $V_{\rm eff}$ is obtained as
\begin{eqnarray}
V_{\rm eff}(\phi)=\frac{1}{(1+\xi f(\phi))^2}\left(\lambda^2f(\phi)^2+\mu^4-\frac{3W_0^2}{1+\xi f(\phi)+\frac{3}{2}\xi^2f'(\phi)^2}\right).
\end{eqnarray}
By requiring the condition for the vanishing cosmological constant at the minimum $V_{\rm eff}(\phi_{\rm min})=0$ with $\phi_{\rm min}\sim 0$, the parameter $\mu$  is written by the other ones and then $V_{\rm eff}$ is written as
\begin{eqnarray}
V_{\rm eff}(\phi)=\frac{1}{(1+\xi f(\phi))^2}\left(\lambda^2 f(\phi)^2+3W_0^2\frac{\xi f(\phi)+\xi^2f'(\phi)^2}{1+\xi f(\phi)+\frac{3}{2}\xi^2f'(\phi)^2}\right).\label{CSS_strong}
\end{eqnarray} 
\begin{figure}
\begin{center}
\includegraphics[width=\textwidth]{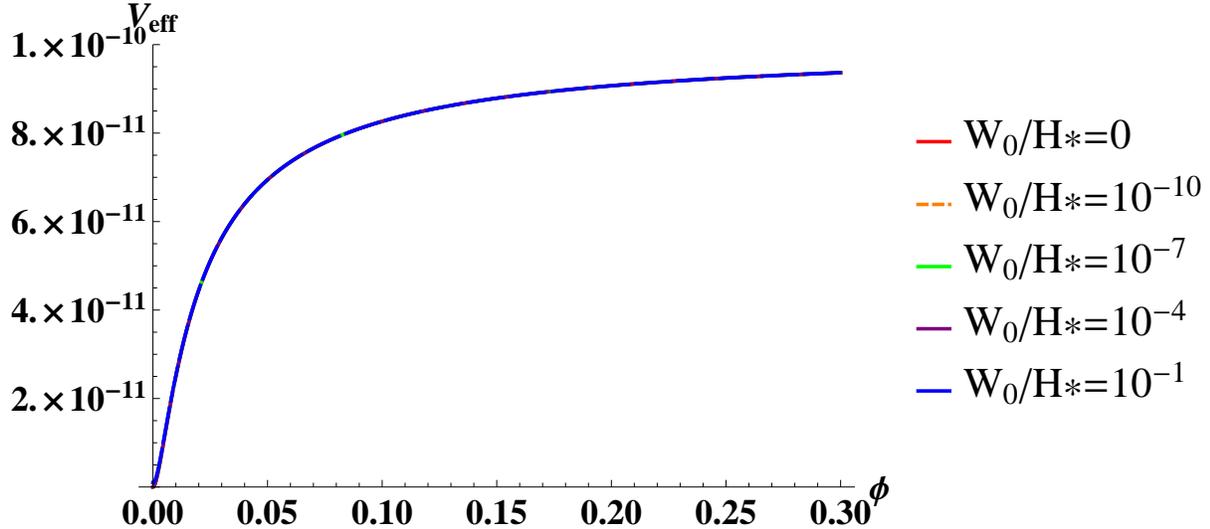}
\caption{The $\phi$-dependences of the scalar potential~(\ref{CSS_strong}) with $\xi=100$ and different values of $W_0$ are shown. In this model, the set of parameters~$(\xi,\lambda)$ are chosen as $(100, 10^{-5})$. The ratio between $W_0$ and the Hubble parameter $H_*\sim\sqrt{\frac{V(N_*=50)}{3}}\sim 5.77\times10^{-6}$ for each line is different from each other ($N_*$ is a number of e-foldings before the inflation end), although those lines are  indistinguishably overlapped.}
\label{CSSstrong_fig}
\end{center}
\end{figure}
For $f(\phi)=\phi$, we show the behaviors of the potential with various SUSY breaking scales in Fig.~\ref{CSSstrong_fig}. In contrast to the case of the minimal K\"ahler model in Sec.\ref{mini}, it is remarkable  that  the deformation of scalar potential is negligibly small even when SUSY breaking scale becomes large. This fact can be easily understood from the arguments in Sec.~\ref{conformal}.
\subsection{The minimal K\"ahler model with a shift symmetry}\label{shift}
We now discuss the chaotic inflation with an F-term potential~\cite{Kawasaki:2000yn,Kallosh:2010ug}. In this model, the K\"ahler potential is given by $K_{\rm shift}$, which is invariant under the following transformation,
\begin{eqnarray}
\Phi\rightarrow \Phi+ic,
\end{eqnarray}
where $c$ is a real constant. The superpotential is given by
\begin{eqnarray}
W=m\Phi S,
\end{eqnarray}
where $m$ is a real constant and $S$ is a chiral multiplet. In general, the scalar component of $S$ plays the role of the sGoldstino during inflation, and a quartic term $|S|^4$ is necessary in the K\"ahler potential to stabilize $S$ during inflation. Therefore, the chaotic inflation in supergravity can be realized with the following K\"ahler potential,
\begin{eqnarray}
K=K_{\rm shift}+|S|^2-\zeta |S|^4,\label{Kchaotic}
\end{eqnarray}  
and the superpotential~(2.31), where $\zeta$ is a real constant. 

As in the other models discussed so far, we add the SUSY breaking sector which has the following K\"ahler and superpotential,
\begin{eqnarray}
\delta K&=&|X|^2-\frac{|X|^4}{\Lambda^2},\\
\delta W&=&W_{\cancel{\rm SUSY}}.
\end{eqnarray} 
When the coefficient of the quartic term $\zeta$ and $1/\Lambda^2$ are sufficiently large, the field values of $S$ and $X$ become small during inflation. Then the scalar potential can be expanded with respect to $S$ and $X$. Therefore we take into account the terms up to $\mathcal{O}(|S|^2,|X|^2,...)$ in the following analysis, and the scalar potential is represented by
\begin{eqnarray}
V&\simeq& e^{\frac{1}{2}(\Phi+\bar{\Phi})^2}\left[m^2|\Phi|^2+(\Phi+\bar{\Phi})^2W_0^2\right]\nonumber\\
&&+m_S^2|S|^2+m_X^2|X|^2+(\alpha_SS+\alpha_XX+m_{S\bar{X}}S\bar{X}+{\rm h.c.}),
\end{eqnarray} 
where $\alpha_{S,X}$ and $m_{S,X}^2$ in the second line are $\Phi$-dependent coefficients which satisfy $m_{I}^2\gg |\alpha_I|$ for $I=S,X$ respectively.

By integrating out the heavy modes $S$ and $X$, we obtain the effective potential during inflation,
\begin{eqnarray}
V_{\rm eff}\simeq e^{\sigma^2}\left[\left(\frac{1}{2}m^2+2W_0^2\right)\sigma^2+\frac{1}{2}m^2\phi^2\right],
\end{eqnarray} 
where we define real fields $\sigma$ and $\phi$ by $\Phi=(\sigma+i\phi)/\sqrt{2}$.

We notice that only the real part $\sigma$ is affected by the SUSY breaking effect. This is a consequence of the shift symmetry in the K\"ahler potential (\ref{Kchaotic}), and then the inflaton $\phi$ is not affected by SUSY breaking terms. After integrating out $\sigma$, we obtain the usual chaotic inflation potential 
\begin{eqnarray}
V_{\rm eff}\simeq \frac{1}{2}m^2\phi^2.
\end{eqnarray} 

In this type of models, SUSY breaking effects do not affect the inflationary trajectory even if the SUSY breaking is larger than the Hubble scale.\footnote{We do not include the back reaction from the VEVs of the scalars other than the inflaton because they become small if the quartic couplings in the K\"ahler potential are sufficiently large. Otherwise, our statement will be changed. We briefly discuss the back reaction effects in Appendix. B. Recently the effects of the back reaction are also studied in Ref.~\cite{Buchmuller:2014pla} by W.~Buchmuller et al, and their results confirm our statements.}
\section{The effective suppression scale of SUSY breaking terms}\label{conformal}
In this section, we interpret the behavior of inflaton potential terms induced by the SUSY breaking, especially their response to a change of the magnitude of SUSY breaking  from the conformal supergravity point of view. The application of conformal supergravity to cosmology was throughly studied in Ref.~\cite{Kallosh:2000ve}. The conformal supergravity description is quite useful to understand the effects of underlying symmetries on cosmology as well as particle physics in supergravity models~\cite{Kallosh:2013oma}. Furthermore, the transformations between different cosmological frames, e.g., the Einstein and the Jordan frames, are easily performed with such a description.\footnote{Recently, inflationary potential in supergravity was reconsidered in terms of the Jordan frame quantities in Ref.~\cite{Kallosh:2014xwa} based on conformal supergravity.} 

As shown in Appendix.~\ref{app}, the structure of the F-term scalar potential is changed depending on the gauge fixing condition of the superconformal symmetry. All the dimensionfull quantities in the chiral matter action~(\ref{confgen}) are related to the field value of the chiral compensator $S_0$, which is fixed by the dilatation gauge condition~(A.6), that is $|S_0|^2=-3/\Omega=e^{K/3}$ in the unit $M_{\rm pl}=1$, corresponding to the Einstein frame. After fixing the superconformal symmetry, the compensator multiplet produces the gravitational corrections to the Lagrangian terms of matter fields, especially to the scalar potential, through their couplings with $S_0$ and the F-term of compensator $\tilde{F}^{S_0}$ in Eq. (\ref{F^S}), those disappear in the decoupling limit of gravity (i.e., the global SUSY limit). Here we call the SUSY breaking terms induced by $\tilde{F}^{S_0}$ ``pure" gravity-mediated ones. On the other hand, there also exist terms involving the F-term itself of chiral multiplet $X$ in the SUSY breaking sector, which also deform the scalar potential thorough the interaction with the compensator. The latter terms are not called "pure" gravity-mediated ones in our terminology.

Let us reconsider the SUSY breaking effects on the inflationary potential discussed in Sec.\ref{models}, those are represented by the potential terms proportional to $W_0^2$ in our setup. The effective potential of the D-term inflation~(\ref{VD}) in the minimal K\"ahler model can be rewritten with the dilatation gauge fixing $\Omega=-3/|S_0|^2$ as,
\begin{eqnarray}
V^{\rm mini}=\frac{g^2\chi^2}{2}-\frac{27W_0^2}{\Omega^3}|\Phi|^2+\frac{g^4\chi^2}{16\pi^2}\left(1+\log \frac{-3\sqrt{3}\lambda|\Phi|^2}{g^2\chi\Omega^{3/2}}\right).\label{VDc}
\end{eqnarray}
We notice that the second term in Eq.~(\ref{VDc}) proportional to the factor $\Omega^{-3}$ is the ``pure" gravity-mediated SUSY breaking term. The first and the third terms are originated from the D-term potential and its quantum corrections. They are physically conformal invariant quantities, and then such terms can not couple to the compensator multiplet $\mathcal{S}_0$ directly. Therefore they are not affected by the rescaling factor $\Omega$.

Similarly, the scalar potential in the conformal model~(\ref{CSS_strong}) can also be rewritten with $\Omega$,
\begin{eqnarray}
V^{\rm conf}=\frac{\lambda}{\xi^2}\left(\frac{1+\Omega/3}{2+\Omega/3}\right)^2+\frac{27W_0^2}{\Omega^2}\left(\frac{\xi f(\phi)+\xi^2f'(\phi)^2}{1+\xi f(\phi)+\frac{3}{2}\xi^2f'(\phi)^2}\right)^2.\label{CSSc}
\end{eqnarray}
The second term in Eq.~(\ref{CSSc}) proportional to $\Omega^{-2}$ can be regarded as the SUSY breaking effect induced by the $F$-term itself of $X$. It is worth noting that the difference of the power of $\Omega$ between the ``pure" gravity-mediated term in Eq.~(\ref{VDc}) and the second term in Eq.~(\ref{CSSc}) arises because they have essentially different origins from each other as indicated above (see also Appendix \ref{app}).

Although terms induced by the SUSY breaking are proportional to the negative power of $\Omega$ in both the above models, the sensitivities of them to the SUSY breaking scale are drastically different, as shown in Figs.~\ref{Dterm_fig} and \ref{CSSstrong_fig}.  To understand such a feature, we have to recognize the functional form of $\Omega$ in each model. In the minimal K\"ahler model, $|\Omega|=3\exp (-|\Phi|^2/3)$ decreases as the field value of $|\Phi|$ becomes larger, and therefore the SUSY breaking term is enhanced compared with the other supersymmetric terms in Eq.~(\ref{VDc}). On the other hand, $|\Omega|=3(1+\xi f(\phi))$ in the conformal model increases as the field value of the inflaton $\phi$ getting larger. Then the SUSY breaking term is suppressed at the large field value of $\phi$ in Eq.~(\ref{CSSc}).  

In any case, the field-dependent factor $\Omega$ can be interpret as the rescaling factor of the physical dimensionfull quantities of the chiral matter action. We show the simple explanation of $\Omega$ playing a role of the rescaling factor in Appendix~\ref{app}, and here we just summarize it. In the conformal supergravity framework, there are no dimensionfull parameters in the action~(\ref{confgen}), and the dimensionful quantities arise after fixing the dilatation symmetry. Then, in the Poincar\'e supergravity in the Einstein-frame, dimensionfull quantities appear as a consequence of rescaling fields by the dilatation-fixing scale determined as shown in Eq.~(\ref{Dgc}). The effects of the rescaling are governed by the factor $\Omega$ depending on the couplings of matter fields with the compensator multiplet. Therefore, the gravity-mediated supersymmetry breaking terms are suppressed or enhanced by such a rescaling factor. Furthermore, because the $F$-component itself of chiral compensator can induce SUSY breaking terms in general, especially at the SUSY breaking Minkowski miminum of the supergravity potential, they are distinctive from the other SUSY breaking terms with the different power of $\Omega$ in their factors.

We also have to mention the terms which are not suppressed by the rescaling factor. The first and the third term in Eq.~(\ref{VDc}) arise from the gauge interaction which possesses a physical conformal symmetry at the classical level.\footnote{We can also say that the gauge multiplets have their Weyl weight compatible with their canonical dimensions, and therefore they do not interact with the compensator $S_0$.} On the other hand, the first term in Eq.~(\ref{CSSc}) is indeed suppressed by the factor $\mathcal{O}(1/\Omega)$ for a large $\Omega$, however, the suppression is canceled by the same power of $\Omega$ in the numerator and such a mechanism leads to the universal attractor behavior ~\cite{Kallosh:2013tua}, that is a consequence of the special form of K\"ahler potential~(\ref{KCSS}) in this model. 

Finally, the chaotic inflation model discussed in Sec.~\ref{shift} does not suffer from the SUSY breaking effects. This fact is by virtue of the shift symmetry responsible for the flatness of (supersymmetric) inflaton potential, which also forbids the existence of the ``pure" gravity-mediated SUSY breaking terms of the inflaton.\footnote{
The shift symmetric structure generally appears for the axions in string theory, where the axion mass is generated by some non-perturbative superpotential at the leading order. Such a fact is important for, e.g., the realization of multi-natural inflation in supergravity~\cite{Czerny:2014xja}. }
It is remarkable that the shift symmetric structure of the K\"ahler potential assures that the rescaling factor $\Omega$ does not depend on the field shifted by the symmetry transformation. In such a model, it is described as
\begin{eqnarray}
|\Omega|=3\exp\left(-\frac{K(\Phi+\bar{\Phi})}{3}\right).
\end{eqnarray}  
If the inflaton field is identified as ${\rm Im}~\Phi$, the above mentioned rescaling factor does not decrease or increase even when the field value of inflaton varies. Therefore, the gravity-mediated SUSY-breaking effects are not enhanced even if they exist.\footnote{In our setup, we do not include the couplings between $X$ and $\Phi$, such as $c |X|^2|\Phi|^2$ in the K\"ahler potential, which break the shift symmetry. If such a coupling exists, the gravity-mediated scalar mass of inflaton is induced.} 
\section{Implications from BICEP2 results}
Recently, the BICEP2 collaboration reported the non-zero value of tensor-to-scalar ratio $r$. Although there is a tension between the results from the BICEP2 and Planck experiments, it is important to discuss the class of models which are compatible with the BICEP2 result. Therefore, in this section, we focus on such a class of models and discuss the behavior of the SUSY breaking effects in those models.
The observed value of the tensor-to-scalar ratio is 
\begin{eqnarray}
r=0.16^{+0.06}_{-0.05}\label{rB},
\end{eqnarray}
after subtracting the best available estimate for a foreground dust.

Such a large tensor-to-scalar ratio can be realized in the models discussed in Sec.~\ref{CSS} and \ref{shift}. The usual chaotic inflation discussed in Sec.~\ref{shift} predicts \begin{eqnarray}
r=\frac{8}{N_*}=0.16\quad ({\rm for}\ N_*=50),
\end{eqnarray}
which is compatible with the observed value~(\ref{rB}).

\begin{figure}
\begin{center}
\includegraphics[width=\textwidth]{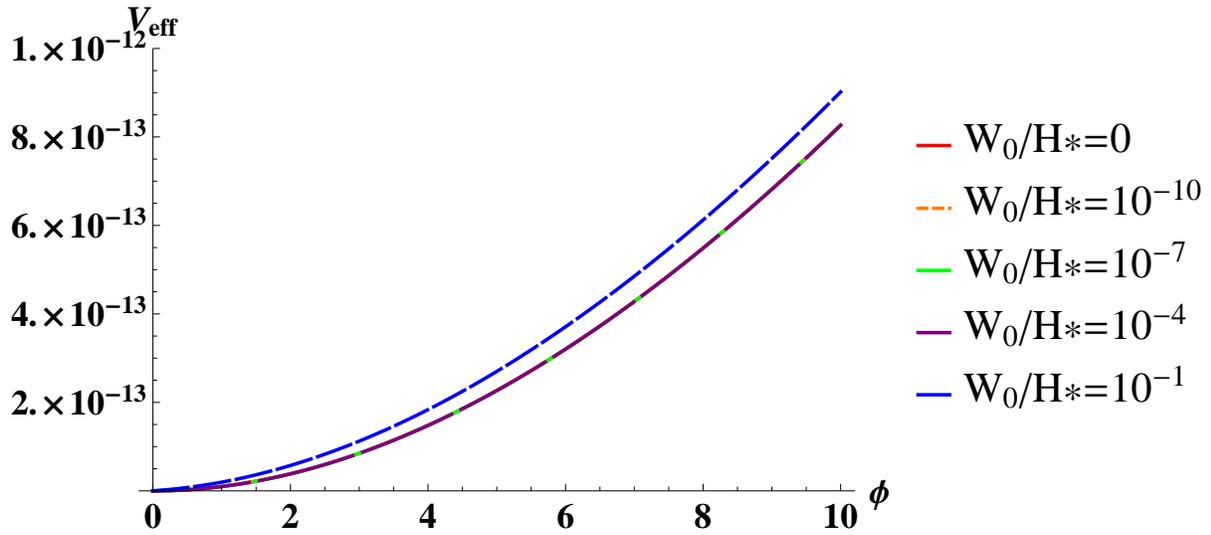}
\caption{The behaviors of the scalar potential~(\ref{CSS_strong}) and various values of $W_0$ are shown. The set of model parameters~$(\xi,\lambda)$ is chosen as $(0.01,10^{-5})$. The SUSY breaking scale $W_0 \sim m_{3/2}$ is normalized by the Hubble parameter~$H_*\sim 5.77\times10^{-6}$ at $N_*=50$, which is the number of e-foldings before the end of inflation. Except for the case with $W_0/H_*=0.1$, all the lines are indistinguishably degenerated.}
\label{CSSweak_fig}
\end{center}
\end{figure}

The model discussed in Sec.~\ref{CSS} can realize the similar value to Eq.~(\ref{rB}) if the constant $\xi$ is much smaller than 1, because a shift-symmetry for the inflaton $\phi$ is restored in the limit of $\xi\rightarrow 0$. With a small but non-zero $\xi$, the inflaton potential is affected by the SUSY breaking effects through the small shift symmetry breaking term in the K\"ahler potential.  In Fig.~\ref{CSSweak_fig}, we show the scalar potential with the SUSY breaking effects in the case of $\xi \ll1$. We recognize that the SUSY breaking effects change the shape of inflaton potential in this case. In the case of $\xi>1$ which predicts the almost vanishing $r$, the SUSY breaking effects are sufficiently suppressed and the scalar potential is not affected even if the SUSY breaking scale is comparable with the Hubble scale during the inflation. However, in the former case with $\xi\ll1$ required by BICEP2, the absolute value of $\Omega$ discussed in the previous section does not become large enough during the inflation, which makes the inflaton potential more sensitive to the SUSY breaking than the case of $\xi>1$.\footnote{It is remarkable that although the SUSY breaking effects deform the potential as shown in Fig.~\ref{CSSweak_fig}, the deformation does not occur drastically compared with the case of the minimal K\"ahler model. That is due to the functional form of the rescaling factor $\Omega$. } Even in the model with  $K_{\rm shift}$ studied in Sec. 2.3, a similar deformation of the potential can occur in the case that its shift symmetry is broken.\footnote{The effects of the shift symmetry breaking are studied in Refs.~\cite{Kallosh:2010ug,Li:2013nfa,Harigaya:2014qza}}

In the case that the inflaton potential is sensitive to the SUSY breaking, the breaking scale would be extracted model by model from the estimation of its effects on the predicted values of the cosmological parameters which are compared with the observational results, where the effective description proposed in Sec. 3 will be useful. That is also important from the viewpoint of particle phenomenology. Sparticles have not been discovered at the LHC experiments, while the discovery of Higgs boson with its mass 126GeV may imply that the SUSY breaking scale is very high. If it is the case, it is extremely difficult to seek the evidence for SUSY at the collider experiments, however, the effects of sparticles (or SUSY breaking) may be proved by the cosmological observations based on the above considerations.  
\section{Conclusion}
In this paper, we have discussed the SUSY breaking effects on typical inflation mechanisms. As shown in Sec.~2, the model with $K_{\rm min}$ is susceptible to the SUSY breaking effects, while the models with $K_{\rm conf}$ and $K_{\rm shift}$ are not so. 

Such a difference can be easily understood by noticing the field-dependent rescaling factor of the gravity-mediated SUSY breaking terms based on the conformal supergravity framework. In both of the latter two models, the terms induced by the SUSY breaking are suppressed by the rescaling factor $\Omega=-3\exp (-K/3)$. During the inflationary era, $|\Omega|=3\exp (-K/3)$ is small with $K_{\rm min}$, while it becomes large (totally unchanged) with $K_{\rm conf}$ ($K_{\rm shift}$), that is the essential difference between these typical inflation models.
As discussed in Sec. \ref{conformal}, the physical dimensionful parameters are rescaled by the compensator, accompanied by the field-dependent factor $\Omega$, and the behavior of the SUSY breaking terms is governed by the structure of the rescaling factor in the Einstein frame.

In the same way, we can understand that the so-called $\eta$ problem in the large field regime is caused by the enhancement of the terms due to the rescaling factor.\footnote{In Ref.~\cite{Kallosh:2014rga}, the related observation was performed in studying the inflationary attractor behavior. } Understanding in this manner would be useful to construct the models of inflation and particle cosmology.
For example, the recent result from BICEP2 supports the inflation mechanisms with a large field value of inflaton, such as the chaotic inflation. In the supergravity framework, such large-field models lead to a decreasing or increasing $|\Omega|$ in general, which spoils the flatness of the inflaton potential when the SUSY breaking is incorporated. Then, certain symmetries of the K\"ahler potential terms restricting the structure of $\Omega$ will be important to avoid such a problem. As we have shown, for example, the shift symmetry in the chaotic inflation in supergravity plays essential roles to ensure the flatness of scalar potential even with the SUSY breaking, because the symmetry forbids the decreasing feature of the factor $\Omega$ and restricts the deformation of the potential caused by SUSY breaking effects. Therefore, it would be important to seek the origin of such a symmetry in some UV completions of these inflation models.

\subsection*{Acknowledgements}
H.A. was supported in part by the Grant-in-Aid for Scientific Research No. 25800158 from the Ministry of Education, Culture, Sports, Science and Technology (MEXT) in Japan.  Y.Y. was supported in part by JSPS Research Fellowships for Young Scientists No. 26-4236 and a Grant for Excellent Graduate Schools from the MEXT in Japan.

\appendix
\section{A general description of F-term potential with conformal supergravity}\label{app}
We briefly review the structure of the F-term scalar potential in terms of the conformal supergravity.\footnote{For more details, see, e.g., Ref.~\cite{Kugo:1982mr} and references there in.} Here, we explicitly denote the Planck scale $M_{\rm pl}$ to clarify mass dimensions of each quantities. In the framework of ${\cal N}=1$ conformal supergravity with the spacetime-dimension four, an unphysical multiplet is introduced to fix the extra symmetries for obtaining the Poincar\'e supergravity, which is called the chiral compensator multiplet denoted by ${\cal S}_0$ in this paper. Then the general action for physical chiral multiplets $\Phi^I$ is written by
\begin{eqnarray}
\left[{\cal S}_0 \bar{\cal S}_0\Omega(\Phi^I,\bar{\Phi}^{\bar{J}})\right]_D+\left[{\cal S}_0^3{\cal W}(\Phi^I)\right]_F,\label{confgen}
\end{eqnarray}
where $[\cdots]_{D,F}$ denote the D- and F-term superconformal density formulae, respectively, and $\bar{\Phi}^{\bar{J}}$ is the Hermitian conjugate of $\Phi^J$. The action~(\ref{confgen}) is determined by the Weyl and chiral weights $(w,n)$ assigned to $S_0$ with $(1,1)$ and to $\Phi^I$ with $(0,0)$, respectively. We have to stress that the action~(\ref{confgen}) does not contain any dimensionfull quantities due to the superconformal symmetry, and therefore, all the components of each superconformal multiplets are dimensionless.

We focus on the F-term potential contained in the action~(\ref{confgen}), that is written as 
\begin{eqnarray}
-V_{F}=S_0\bar{S}_0\Omega_{I\bar{J}}F^I\bar{F}^{\bar{J}}+\Omega F^{S_0}\bar{F}^{\bar{S}_0}+\left(S_0\Omega_{\bar{J}}F^{S_0}\bar{F}^{\bar{J}}+S^3_0{\cal W}_IF^I+3F^{S_0}S_0^2{\cal W}+{\rm h.c.}\right),
\end{eqnarray}
where $S_0$ and $F^{S_0}$ are the scalar and F-component of $\mathcal{S}_0$, respectively, and $F^I$ is the F-component of $\Phi^I$. Note that $\Omega$ and ${\cal W}$ (without the arguments) are not the functions of chiral multiplets $\Phi^I$ themselves but of the scalar components of $\Phi^I$. First, we integrate out $F^{S_0}$ and its conjugate, and obtain the following expression,
\begin{eqnarray}
-V_F=S_0\bar{S}_0\Omega_{I\bar{J}}F^I \bar{F}^{\bar{J}}+\left(S_0^3\mathcal{W}_IF^I+{\rm h.c.} \right)-\Omega \tilde{F}^{S_0}\tilde{\bar{F}}^{\bar{S}_0},
\end{eqnarray}
where
\begin{eqnarray}
\tilde{F}^{S_0}=-\frac{1}{\Omega}\left(\bar{S}_0\Omega_IF^I+3\bar{S}_0^2\bar{\cal W}\right).\label{F^S}
\end{eqnarray}
 We can also integrate out $F^I$ and its conjugate, and find the following scalar potential $V_F$,
 \begin{eqnarray}
V_F=|S_0|^4\mathcal{M}^{I\bar{J}}\left({\cal W}_I-\frac{3\Omega_I}{\Omega}{\cal W}\right)\left(\bar{\cal W}_{\bar{J}}-\frac{3\Omega_{\bar{J}}}{\Omega}\bar{\cal W}\right)+\frac{9|S_0|^4}{\Omega}|{\cal W}|^2,\label{Vc}
\end{eqnarray}
where $\mathcal{M}_{I\bar{J}}\equiv\Omega_{I\bar{J}}-\Omega^{-1}\Omega_I\Omega_{\bar{J}}$, and $\mathcal{M}^{I\bar{J}}$ is the inverse of $\mathcal{M}_{I\bar{J}}$.

 To obtain the Poincar\'e supergravity action in the Einstein frame, we choose the following gauge-fixing conditions,
\begin{eqnarray}
S_0\bar{S}_0\Omega=-3M_{\rm pl}^2,\label{Dgc}
\end{eqnarray} 
and
\begin{eqnarray}
S_0=\bar{S}_0,
\end{eqnarray}
which fix the extra (dilatation and $U(1)$) gauge symmetries those are absent in the Poincar\'e supergravity.
Then the conditions can be solve in terms of $S_0$ and $\bar{S}_0$ as follows,
\begin{eqnarray}
S_0=\bar{S}_0=\left(-\frac{3}{\Omega}\right)^{\frac{1}{2}}M_{\rm pl}.\label{gauge}
\end{eqnarray}
We note that the mass dimension is introduced through these procedure, and also recognize that the canonical mass dimension in the physical theory corresponds to the Weyl weight because the Weyl weight of $S_0$ corresponds to the mass dimension of the right-hand side of (\ref{gauge}). 

Substituting Eq.~(\ref{gauge}) into Eq.~(\ref{Vc}), the scalar potential $V_F$ is expressed by
\begin{eqnarray}
V_F=M_{\rm pl}^4\frac{9}{\Omega^2}\mathcal{M}^{I\bar{J}}\left({\cal W}_I-\frac{3\Omega_I}{\Omega}{\cal W}\right)\left(\bar{\cal W}_{\bar{J}}-\frac{3\Omega_{\bar{J}}}{\Omega}\bar{\cal W}\right)+M_{\rm pl}^4\frac{81}{\Omega^3}|{\cal W}|^2.\label{VC}
\end{eqnarray}

Although the Planck mass is introduced to the action, the superpotential ${\cal W}$ and $\phi^I$ which is the scalar components of $\Phi^I$ are dimensionless. That is because $\phi^I$ and $\bar{\phi}^{\bar{J}}$ have zero Weyl weights respectively. Their canonical dimensions are adjusted to the physical ones by the couplings with the compensator $S_0$, and that is also true for the superpotential ${\cal W}$. Therefore such dimensionfull quantities are associated with the factor $\left(-\frac{3}{\Omega}\right)^{\frac{1}{2}}$ in Eq.(\ref{gauge}).\footnote{Although we only consider chiral multiplets here, these statements are applicable to all the other types of multiplets which do not have their Weyl weights not compatible with their canonical dimensions.} Then, we redefine $\phi^I$ as well as ${\cal W}$ by the following relation: \footnote{We replace $\phi^I$ to $\tilde{\phi}^I$ to clarify the relation between the quantities in conformal SUGRA and physical ones, however we use the unit $M_{\rm pl}=1$ in the other sections. Then all the expressions are not altered there by the replacement. Therefore, we don't distinguish $\phi^I$ with $\tilde{\phi}^I$, except for ones in this Appendix.}
\begin{eqnarray}
\phi^I=\tilde{\phi}^I/M_{\rm pl},\\
{\cal W}(\phi)=W(\tilde{\phi}^I)/M_{\rm pl}^3.
\end{eqnarray}     
After these procedures, the action~(\ref{VC}) is rewritten as
\begin{eqnarray}
V_F=\frac{9}{\Omega^2}\mathcal{M}^{I\bar{J}}\left(W_I-\frac{3\Omega_I}{\Omega M_{\rm pl}^2}W\right)\left(\bar{W}_{\bar{J}}-\frac{3\Omega_{\bar{J}}}{\Omega M_{\rm pl}^2}\bar{W}\right)+\frac{81}{\Omega^3M_{\rm pl}^2}|W|^2.\label{FP}
\end{eqnarray}

The physical K\"ahler potential $K$ is related to $\Omega$ as 
\begin{eqnarray}
\Omega=-3e^{-K/3M_{\rm pl}^2}.\label{KO}
\end{eqnarray}
One can easily confirm that the F-term scalar potential~(\ref{VC}) is equivalent to the expression in Eq.~(\ref{VE}) with Eq.~(\ref{KO}).

The $\Omega$ dependence of each terms in the conformal model is obvious because $\Omega$ is given by the polynomial of the scalar fields.
On the other hand, in the minimal K\"ahler model~(\ref{VDc}),
\begin{eqnarray}
K=\sum_I|\tilde{\phi}^I|^2,
\end{eqnarray}
the matrix ${\cal M}_{I\bar{J}}$ is
\begin{eqnarray}
{\cal M}_{I\bar{J}}&=&-\frac{1}{3}\Omega\left(\delta_{I\bar{J}}-\frac{1}{3M_{\rm pl}^2}\Phi^J\bar{\Phi}^{\bar{I}}\right)\\
&\equiv&-\frac{1}{3}\Omega \tilde{\cal M}_{I\bar{J}}.
\end{eqnarray}
Then the scalar potential (\ref{VC}) can be rewritten as
\begin{eqnarray}
V_F=-\frac{27}{\Omega^3}\left[ \tilde{\cal M}^{I\bar{J}}(W_I+\frac{\bar{\Phi}^{\bar{I}}}{M_{\rm pl}^2}W)(\bar{W}_{\bar{J}}+\frac{\Phi^J}{M_{\rm pl}^2}\bar{W})-3|W|^2\right],
\end{eqnarray}
where $\tilde{\cal M}^{I\bar{J}}$ is an inverse of $\tilde{\cal M}_{I\bar{J}}$. Then the SUSY breaking term in the minimal K\"ahler model can be easily found as
\begin{eqnarray}
-\frac{27}{\Omega^3}\frac{\bar{\Phi}}{M_{\rm pl}^2}W\times\frac{\Phi}{M_{\rm pl}^2}\bar{W} \sim-\frac{27W_0^2}{\Omega^3M_{\rm pl}^4}|\Phi|^2.
\end{eqnarray} 
\section{Estimation of the back reaction effects}
In \ref{CSS} and \ref{shift}, an additional multiplet $S$ and a SUSY breaking multiplet $X$ are required to have quartic terms in the K\"ahler potential, and the quartic coupling constants $1/\Lambda^2$ and $\zeta$ in~Eqs.(\ref{KX}),~(\ref{Ocss}) and (\ref{Kchaotic}) should be sufficiently large. Here we quantify how large $\zeta$ is required in order the back reaction effects we ignore in Secs.~\ref{CSS} and \ref{shift} to be sufficiently small. 

First, we discuss the case in Sec.~\ref{shift}. In this case, the terms of $S$, $X$ and $\sigma$ are written as,
\begin{align}
\delta V_{\rm shift}\sim& -2W_0\mu^2(X+\bar{X})+\sqrt{2}m\phi W_0(i\bar{S}-iS)+m_X^2|X|^2+m_S^2|S|^2\nonumber\\
&+\frac{1}{2}m_\sigma^2\sigma^2+\frac{m\phi\mu^2}{\sqrt{2}}(-iS\bar{X}+i\bar{S}X),
\end{align}
where
\begin{align}
m_X^2\equiv& \frac{4\mu^4}{\Lambda^2}-2W_0^2+\frac{1}{2}m^2\phi^2,\\
m_S^2\equiv& 2\zeta m^2\phi^2+\mu^4-2W_0^2,\\
m_\sigma^2\equiv& m^2\phi^2+2\mu^4-2W_0^2,
\end{align}
and we neglect the other smaller terms. We can evaluate the minimum of the potential during inflation and obtain the field values of $S$, $X$, and $\sigma$ at the minimum during inflation as,
\begin{align}
X=&\frac{2W_0\mu^2(m_S^2+m^2\phi^2/2)}{m_S^2m_X^2-\mu^4m^2\phi^2/2},\label{vevX}\\
S=&\frac{-i\sqrt{2}m\phi W_0(m_X^2+\mu^4)}{m_S^2m_X^2-\mu^4m^2\phi^2/2},\label{vevS}\\
\sigma=&0\label{vevs}.
\end{align}
By substituting Eqs.~(\ref{vevX})-(\ref{vevs}), the expectation value of the back reaction potential at the minimum, denoted by $\delta V_{\rm shift}|_{\rm min}$, and the ratio between $\delta V_{\rm shift}|_{\rm min}$ and $3H^2=\frac{1}{2}m^2\phi^2$ are estimated as,
\begin{eqnarray}
\frac{\delta V_{\rm shift}|_{\rm min}}{3H^2}\sim \Biggl \{ \begin{array}{ll}
-\frac{W_0^2}{\zeta H^2}& ({\rm for }\ W_0/H\ll1 )\\
-\frac{\Lambda^2W_0^2}{3H^2}-\frac{W_0^2}{3\zeta H^2}& ({\rm for }\ W_0/H\gg1)
\end{array} ,\label{rshift}
\end{eqnarray}
where $W_0/H=W_0/\sqrt{m^2\phi^2/6}$. As we find from Eq.~(\ref{rshift}), the back reaction can be negligible for $\zeta\gg 1$ and $\Lambda\ll 1$ as we mentioned in Footnote~4. For example, if $\zeta,1/\Lambda^2>10W_0^2/(3H^2)$ and $W_0/H\gg1$, the back reaction from $S$ and $X$ is of $\mathcal{O}(0.1 H^2)$.

Next, we discuss the case of the conformal model in Sec.~\ref{CSS} in the same way. For simplicity, we consider the parameter region $W_0/H<0.1$, which we discuss in Sec.~\ref{CSS}. We show the terms involving the relevant fields $S$ and $X$, denoted by $\delta V_{\rm conf}$,
\begin{align}
\delta V_{\rm conf}\sim& \left(-\frac{W_0\lambda f}{\Omega^3}S-\frac{\lambda^2f^2}{\Omega^3}S^2+{\rm h.c.}\right)+\left(\frac{6\zeta\lambda^2f^2}{\Omega^3}+\frac{2W_0^2}{\Omega}\right)|S|^2\nonumber\\
&-\left(\left(\frac{W_0}{\xi \Omega}\right)^2X+\frac{3W_0^2}{\Omega^4}X^2+{\rm h.c.}\right)+\left(2\frac{\lambda^2f^2}{\Omega^2}+3\left(\frac{4}{\Lambda^2}-\frac{2}{3\Omega}\right)W_0^2\right)|X|^2,\nonumber\\
\end{align}  
where we only show the leading terms of $W_0$.
After integrating out $S$ and $\bar{S}$, we can estimate the following back reaction term as,
\begin{align}
\delta V_{\rm conf}|_{\rm min}\sim -\frac{3}{6\zeta -\frac{2}{3}}\frac{|W_0|^2}{\Omega^3}.
\end{align} 
Let us compare the inflationary potential $V_{\rm eff}$ in Eq.~(\ref{CSS_strong}) and $\delta V_{\rm conf}|_{\rm min}$ as in the previous case. Their ratio is evaluated as,
\begin{align}
\frac{\delta V_{\rm conf}|_{\rm min}}{V_{\rm eff}}\sim \frac{W_0^2}{(6\zeta-2/3)\Omega^3H^2}.\label{rconf}
\end{align} 
As we find in Eq.~(\ref{rconf}), the back reaction is not only suppressed if $\zeta$ is sufficiently large but also suppressed if the rescaling factor $\Omega$ is large enough. Therefore, for the parameter space $r<0.1$ for example, which we discuss in Sec.~\ref{CSS}, we can neglect the back reaction effect even with $\zeta\sim 0$. Note, however, that in this concrete example the mass eigenvalues of $S$ and $\bar{S}$ during inflation become tachyonic for $\zeta<1/9$, which can be avoided by requiring $\zeta>1/9$ without spoiling the smallness of the back reaction~(\ref{rconf}).

We should remark that, in both of cases, the vacua after inflation become nontrivial ones when the SUSY breaking scale is much larger than the scale of inflation. In this paper, we don't discuss the vacua after the inflation because they highly depend on the detailed setup of models. In any case, we conclude that the flatness of the inflaton potential is preserved for sufficiently large values of $\zeta$ and $1/\Lambda^2$.

\end{document}